\documentclass[12pt,a4paper]{article}
\usepackage{amsmath,ulem}
\usepackage[font=small,labelfont=bf]{caption}
\usepackage[top=1.2in, bottom=1.2in, left=1.1in, right=1.1in]{geometry}

\DeclareCaptionStyle{italic}[justification=centering]{labelfont={bf},textfont={it},labelsep=colon}
\DeclareMathOperator*{\argmax}{arg\,max}
\usepackage{graphicx,psfrag,epsf}
\usepackage{enumerate}
\usepackage{natbib}
\usepackage{url} 
\usepackage{booktabs, subfig, bm, paralist,mathpazo,tikz,todonotes,longtable,microtype,dsfont,rotating} 
\usepackage[pdftex,colorlinks=true]{hyperref}
\definecolor{darkblue}{rgb}{0,0,.6}
\hypersetup{citecolor=darkblue,linkcolor=darkblue,urlcolor=darkblue}

\newcommand{\blind}{0}
\newcommand{\X}{\mathcal{X}}

\graphicspath{{plots/}}
\DeclareMathOperator*{\argmin}{\arg\!\min}
\newsavebox\CBox
\def\textBF#1{\sbox\CBox{#1}\resizebox{\wd\CBox}{\ht\CBox}{\textbf{#1}}}

\definecolor{a0}{rgb}{0.0, 0.5, 0.0}
\definecolor{bistre}{rgb}{0.24, 0.17, 0.12}
\definecolor{amethyst}{rgb}{0.6, 0.4, 0.8}
\definecolor{blue-violet}{rgb}{0.54, 0.17, 0.89}
\definecolor{Rcolor}{RGB}{150,160,190}
\definecolor{blush}{rgb}{0.87, 0.36, 0.51}
\definecolor{brightturquoise}{rgb}{0.03, 0.91, 0.87}
\definecolor{burntorange}{rgb}{0.8, 0.33, 0.0}
\date{\today}
\AtBeginDocument{}

\begin{document}

\def\spacingset#1{\renewcommand{\baselinestretch}
{#1}\small\normalsize} \spacingset{1}

\if0\blind
{
  \title{\bf Uncovering predictability in the evolution of the WTI oil futures curve}
  \author{Fearghal Kearney\thanks{Corresponding author: Queen's Management School, Queen's University Belfast, BT9 5EE, UK. Telephone: +44 28 9097 4795. Email: f.kearney@qub.ac.uk}
  \hspace{.2cm}\\
    Queen's Management School \\
    Queen's University Belfast \\
\\
    Han Lin Shang\thanks{Research School of Finance, Actuarial Studies and Statistics, Level 4, Building 26C, Australian National University, Kingsley Street, Canberra ACT 2601, Australia. Telephone: +61 2 6125 0535. Email: hanlin.shang@anu.edu.au} \\
    Research School of Finance, Actuarial Studies and Statistics \\
    Australian National University
    }
  \maketitle
} \fi

\if1\blind
{
    \title{\bf Uncovering predictability in the evolution of the WTI oil futures curve}
  \maketitle
} \fi

\bigskip

\begin{abstract}
Accurately forecasting the price of oil, the world's most actively traded commodity, is of great importance to both academics and practitioners. We contribute by proposing a functional time series based method to model and forecast oil futures. Our approach boasts a number of theoretical and practical advantages including effectively exploiting underlying process dynamics missed by classical discrete approaches. We evaluate the finite-sample performance against established benchmarks using a model confidence set test. A realistic out-of-sample exercise provides strong support for the adoption of our approach with it residing in the superior set of models in all considered instances.
\end{abstract}


%\linenumbers

\newpage
\spacingset{1.48}

\section{Introduction}\label{sec:1}

Traditionally, statistical frameworks adopted in both the academic literature and practice address the modelling of low-frequency regularly spaced datasets. However, the recent availability of higher-resolution time series brings with it unique challenges and opportunities \citep{Engle00}. Functional Time Series (FTS) is a convenient modelling and forecasting framework boasting the ability to tackle such problems. \cite{Tsay16} demonstrates that FTS analysis is widely applicable to many areas of big data analytics and business statistics. We take up this mantle and propose a novel exponential smoothing state space FTS framework to address a critical problem in empirical finance, namely the modelling and forecasting of crude oil futures prices.

\cite{AKV13} outline three critical areas for which reliable forecasts of oil prices are essential. First, airlines, automobile firms, utilities and even homeowners make pricing, consumption and investment decisions based on forecasts of oil prices. Second, oil price forecasts are a key component of central bank macroeconomic projections and market practitioners' risk assessments, with more accurate forecasts potentially improving policy responses and risk management strategies. Third, oil price forecasts inform energy usage and carbon emission projections with regulatory policy implications such as fuel excise adjustments and climate change interventions. More specifically, in this paper, we focus on oil futures prices. In general, commodity futures are far more actively traded than their related spot contracts \citep[see, e.g.,][]{NP96}. Outlining the importance of futures markets, \cite{LT04} explicitly identify four functions they serve: efficient price discovery, resource allocation, risk management and financing. Furthermore, energy futures price forecasts can accurately inform the pricing of related securities and projects \citep{KNPP16}.

The use of such financial derivatives in Europe has been widely documented \citep[see, e.g.,][]{COR17}. More specifically for oil derivatives, Brent crude was traditionally regarded as the leading indicator of prices within Europe. However, as the United States (US) recently repealed its 40-year-old ban on exporting WTI (West Texas Intermediate) crude, forecasting WTI crude oil futures has become a vital issue for European audiences. Since the repeal of this ban, there has been a surge in WTI futures trading.\footnote{\textit{Reuters}, ``U.S. oil output surges but growth likely to moderate in 2019", 1 November 2018.} This phenomenon has led to a trend of increased trading volumes for WTI oil futures coupled with reduced trading volumes for the traditionally European Brent oil futures.\footnote{\textit{Financial Times}, ``US oil futures market overtakes London", 17 September 2017.} 

Furthermore, recent production of oil in the European North Sea has seen a structural decline at a time when US WTI crude has observed record production levels. This has led to reports that the firm who set the price of Brent, S\&P Global Platts, is now considering using WTI prices to calculate its latest European index.\footnote{\textit{Bloomberg}, ``U.S. Oil Could Soon Help Set the Brent Crude Price", 6 July 2018} These factors have culminated in WTI arguably becoming the price discovery leader in global crude oil markets and overtaking Brent as the leading oil benchmark in Europe.\footnote{\textit{CME Group}, ``WTI Entering a Renaissance as a Global Benchmark", 17 July 2018.}

Two main classes of models are outlined in the commodity futures modelling literature: discrete latent factor and fundamental macroeconomic-based models. Latent factor approaches to model commodity futures curves include those presented by \cite{CC05, CS08, TS09} and \cite {CLT13}, with a separate strand of literature concluding that fundamentals are important in explaining energy prices \citep[see, e.g.,][]{Kilian09, Singleton13, ABNU16, BP16}. Based on this literature we adopt the fundamental commodity futures forecasting framework of \cite{ABNU16}, as well as the principal component approach of \cite{CS08} as two benchmarks for our proposed FTS model. This aligns with the approach taken by \cite{CDK16} who adopt both frameworks and find no discernible difference between the performance of the latent and fundamental models.

To the best of our knowledge, this paper constitutes the first application of FTS methods to describe, model and forecast commodity futures curves.\footnote{A number of previous empirical studies has benefitted from being cast in a functional data environment. Prominent examples include credit card transactions \citep{Laukaitis08}, online auction price dynamics \citep{WWG08} and electricity price curves \citep{CL17}. More specifically within financial markets, applications span equity index volatility \citep{MSS11}, stock returns \citep{HKR14, HR15, KMZ15, KMRT17, Zhang16}, yield curves \citep{KMR16} and option volatility \citep{LXC16, KMC15, KCM18}.} In the FTS domain, the item of interest is a set of curves, shapes, objects or, more broadly, functional observations. Because trading is focused on specific segments of the oil futures curve, quotes are only observable at discrete expiry points. However, the novel approach that we propose (described in Section 3) seeks to uncover the true underlying curve dynamics that is often missed by traditional techniques. This unearthed underlying process is subsequently projected forward using an exponential smoothing model that has widespread application in business forecasting \citep[see, e.g.,][]{HKOS08, Taylor10}. In an exponential smoothing setting, forecasts are weighted combinations of past observations--in our case, the functional principal component decomposition of daily oil futures curves. The idea is that the most recent observations inform the forecast most heavily, with weights decreasing exponentially for more distant past observations. We evaluate our forecasting models using the data snooping bias-free model confidence set (MCS) procedure of \cite{HLN11}, as also employed by \cite{NS13} and \cite{BM16}. Through this, we ascertain that the FTS approach is superior relative to the benchmark models.

In addition to producing more accurate one-day ahead out-of-sample oil price forecasts, the use of an FTS framework boasts several practical and conceptual advantages. First, the approach effectively captures serial dependence, a prominent attribute of daily oil futures curves. Secondly, it facilitates the modelling of data observed at a small number of unequally spaced points, as well as those with possible non-negligible measurement error, which are common in business statistics \citep[see, e.g.,][]{KMR17}. In our specific context, we follow \cite{CDK16} in focusing on the most actively traded futures contract expiries (CL1--CL9, CL12, and CL18), resulting in unequally spaced data points along the futures curve, which our FTS framework models effectively. Third, the approach facilitates the pricing of non-standard expiry OTC (over-the-counter) contracts, as the output is a complete forecasted futures curve across an essentially infinite number of expiries. Finally, and more practically, given that it relies on latent factors, our approach circumvents the need to a priori identify fundamental factors that model oil futures prices, thus mitigating the risk of omitted variable bias. The paper continues as follows: Section~\ref{sec:2} describes the oil futures dataset, Section~\ref{sec:3} outlines our proposed FTS methodology and forecasting evaluation framework, Section~\ref{sec:4} presents in- and out-of-sample results and Section~\ref{sec:5} concludes the paper.

\section{Oil Futures Data}\label{sec:2}

Crude oil is the world's most heavily traded commodity futures contract \citep[see, e.g.,][]{Sevi14}. More specifically, \cite{BLL18} outline that WTI crude oil has the highest weighting of all futures across the three leading commodity indices compiled by Bloomberg, UBS and Dow Jones. Each WTI futures contract is written on 1,000 barrels of crude oil, whereby a price is agreed today for oil to be delivered and paid for on a specific date in the future. On any given day, a large number of contracts is traded with expiration dates in consecutive months. Generic series can be constructed by rolling from one contract to another, with CL1 representing the oil futures contract with the shortest date to expiry, CL2 the second shortest and so forth. Because of lower trading intensity, liquidity levels are reduced for certain non-standard maturities observed along the futures curve. To alleviate any such concerns, we concentrate only on the most actively traded generic monthly futures in line with \cite{CDK16}. The futures we use are CL1--CL9, CL12 and CL18 obtained from the CME Group. The CME Group is the world's largest marketplace for buying and selling futures contracts. It consists of four separate exchanges: the Chicago Mercantile Exchange, Chicago Board of Trade, New York Mercantile Exchange and Commodity Exchange. Our sample is at a daily frequency from January 2009 to December 2015. To construct the fundamental factor model, daily VIX quotes are downloaded from the Chicago Board of Options Exchange and S\&P 500 (Standard and Poor's 500 index) levels retrieved from Yahoo! Finance, with the Trade Weighted US Dollar Index (USD) and US Economic Policy Uncertainty Index (EcPol) both being obtained from FRED (Federal Reserve Economic Data). Descriptive statistics for our dataset are given in Table~\ref{tab:1}.

\begin{table}[!t]
\begin{center}
\tabcolsep 0.133in
\caption{\small Descriptive Statistics}\label{tab:1}
\begin{tabular}{@{}lrrrrrrr@{}}
\toprule
&	Mean	& Std. Dev.	& Median &	Minimum	& Maximum	& Skewness	& Kurtosis \\
\midrule
CL1      & 81.51   & 19.99  & 86.72   & 33.98  & 113.93  & -0.70 & -0.65 \\
CL2      & 82.07   & 19.48  & 87.41   & 35.78  & 114.43  & -0.70 & -0.66 \\
CL3      & 82.50   & 19.00  & 87.94   & 36.79  & 114.71  & -0.71 & -0.65 \\
CL4      & 82.81   & 18.59  & 88.37   & 37.60  & 114.83  & -0.73 & -0.62 \\
CL5      & 83.05   & 18.20  & 88.77   & 38.45  & 114.82  & -0.74 & -0.59 \\
CL6      & 83.21   & 17.83  & 89.10   & 39.07  & 114.78  & -0.75 & -0.55 \\
CL7      & 83.31   & 17.50  & 89.36   & 39.68  & 114.73  & -0.76 & -0.52 \\
CL8      & 83.39   & 17.18  & 89.52   & 40.24  & 114.58  & -0.77 & -0.48 \\
CL9      & 83.44   & 16.88  & 89.56   & 40.88  & 114.38  & -0.78 & -0.44 \\
CL10     & 83.51   & 16.05  & 89.34   & 42.28  & 113.65  & -0.80 & -0.33 \\
CL11     & 83.32   & 14.59  & 88.21   & 44.41  & 111.71  & -0.83 & -0.15 \\
VIX      & 20.17   & 8.04   & 17.56   & 10.32  & 56.65   & 1.59  & 2.32  \\
S\&P 500 & 1,481.51 & 390.40 & 1,385.14 & 676.53 & 2,130.82 & 0.18  & -1.19 \\
USD      & 77.55   & 6.40   & 75.91   & 68.02  & 94.97   & 1.19  & 0.47  \\
EcPol    & 116.10  & 67.15  & 100.85  & 3.32   & 548.95  & 1.51  & 3.61 \\\bottomrule
\end{tabular}
\end{center}
{\small \textbf{Note:} This table reports the descriptive statistics of CME Group's historical crude oil continuous futures prices. CL1 represent the next consecutive month's futures contract, CL2 is the following month's futures contract and so forth. The sample period is January 2009--December 2015.}
\end{table}

\begin{figure}[!htbp]
\centering
\subfloat[Panel A]
{{\includegraphics[width=8.2cm]{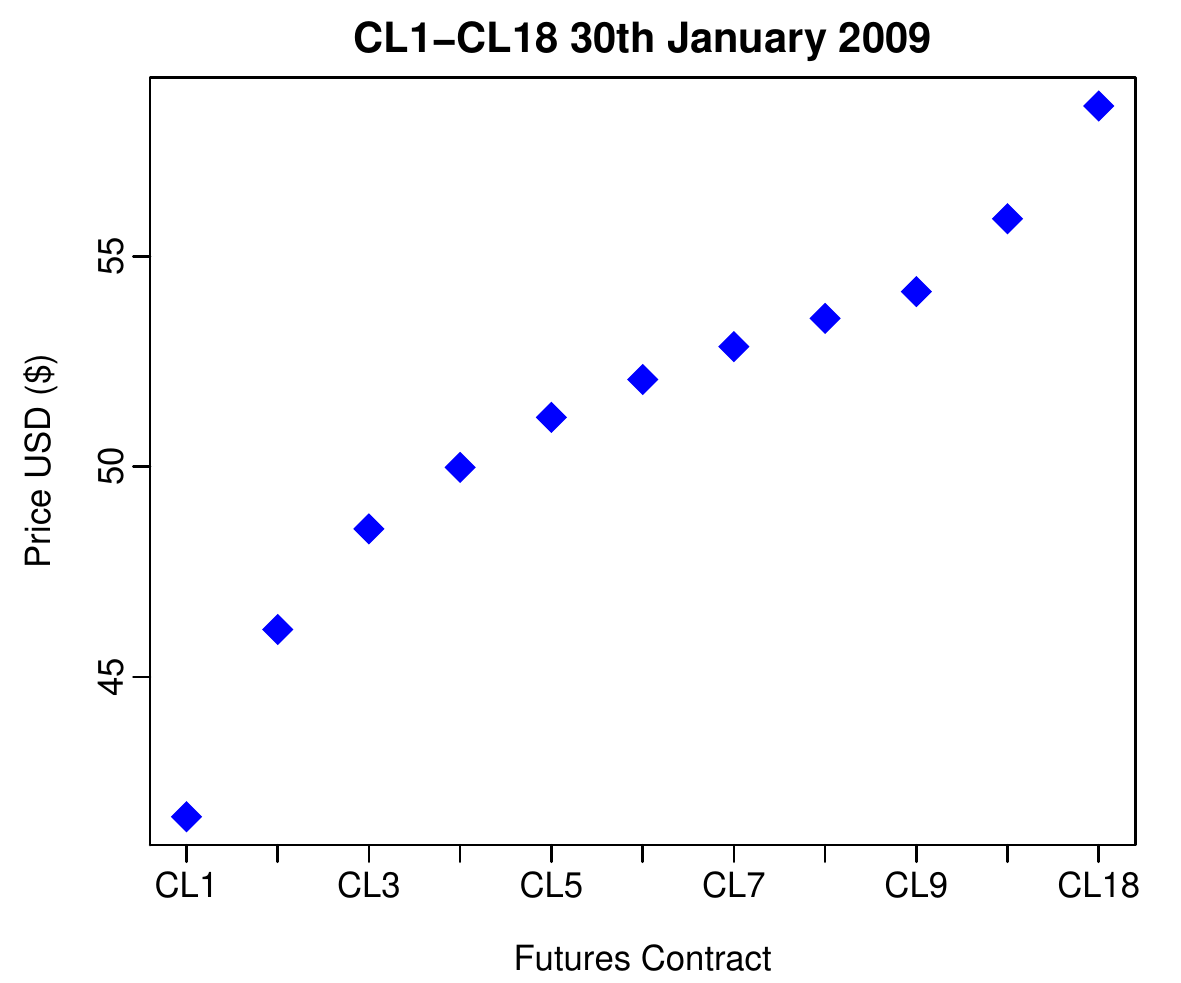}}\label{fig:Fig_1a}}
\qquad
\subfloat[Panel B]
{{\includegraphics[width=8.2cm]{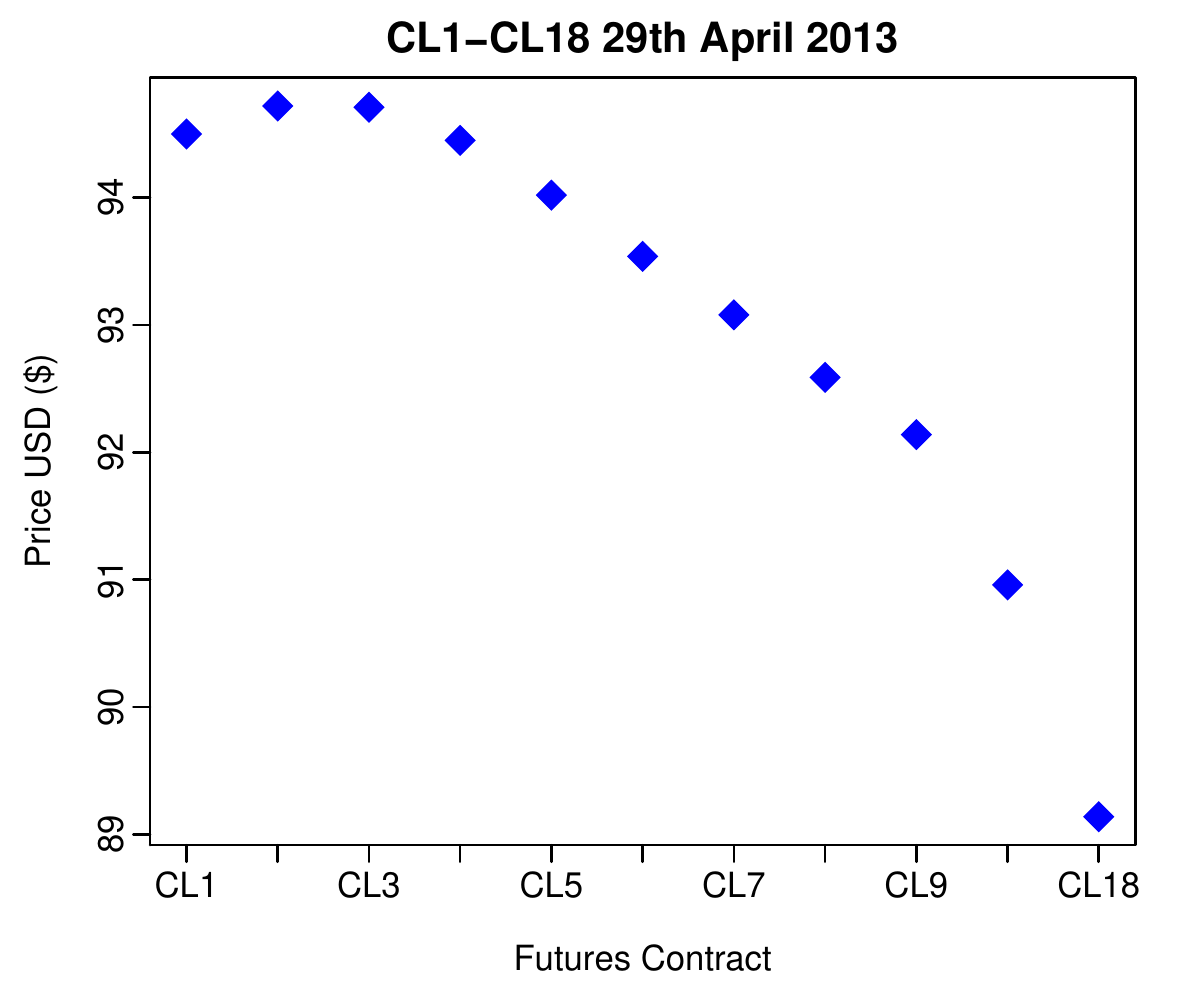}}\label{fig:Fig_1b}}
\caption{\small The crude oil continuous futures price for each of 11 different expiries on the 30 January 2009 (Panel A) and 29 April 2013 (Panel B)}\label{fig:1}
\end{figure}

Figure~\ref{fig:1} presents plots of the relationship between oil futures prices on selected days in 2009 and 2013. In general, the spread between short- and long-expiry futures can be either positive or negative, with the relationship dependent on underlying market forces. On 30 January 2009 (Figure~\ref{fig:Fig_1a}) longer expiries are more expensive than shorter-expiry contracts, which is known as a state of contango. In Figure~\ref{fig:Fig_1b}, the opposite effect is seen, with longer-expiry futures being cheaper, which is known as backwardation. Our flexible FTS approach effectively characterises both states. Furthermore, Figure~\ref{tab:2} presents the evolution of prices for short (1 month), medium (6 months) and long expiry (18 months) futures throughout our sample period. It can be seen that the price spread is not constant over time, with \cite{SS00} previously establishing that such differences between short- and long-maturity contracts provide substantial information about short-term price variations. We next detail our proposed FTS framework to forecast this oil dataset.

\begin{figure}[!htbp]
\centering
\includegraphics[width=11.6cm]{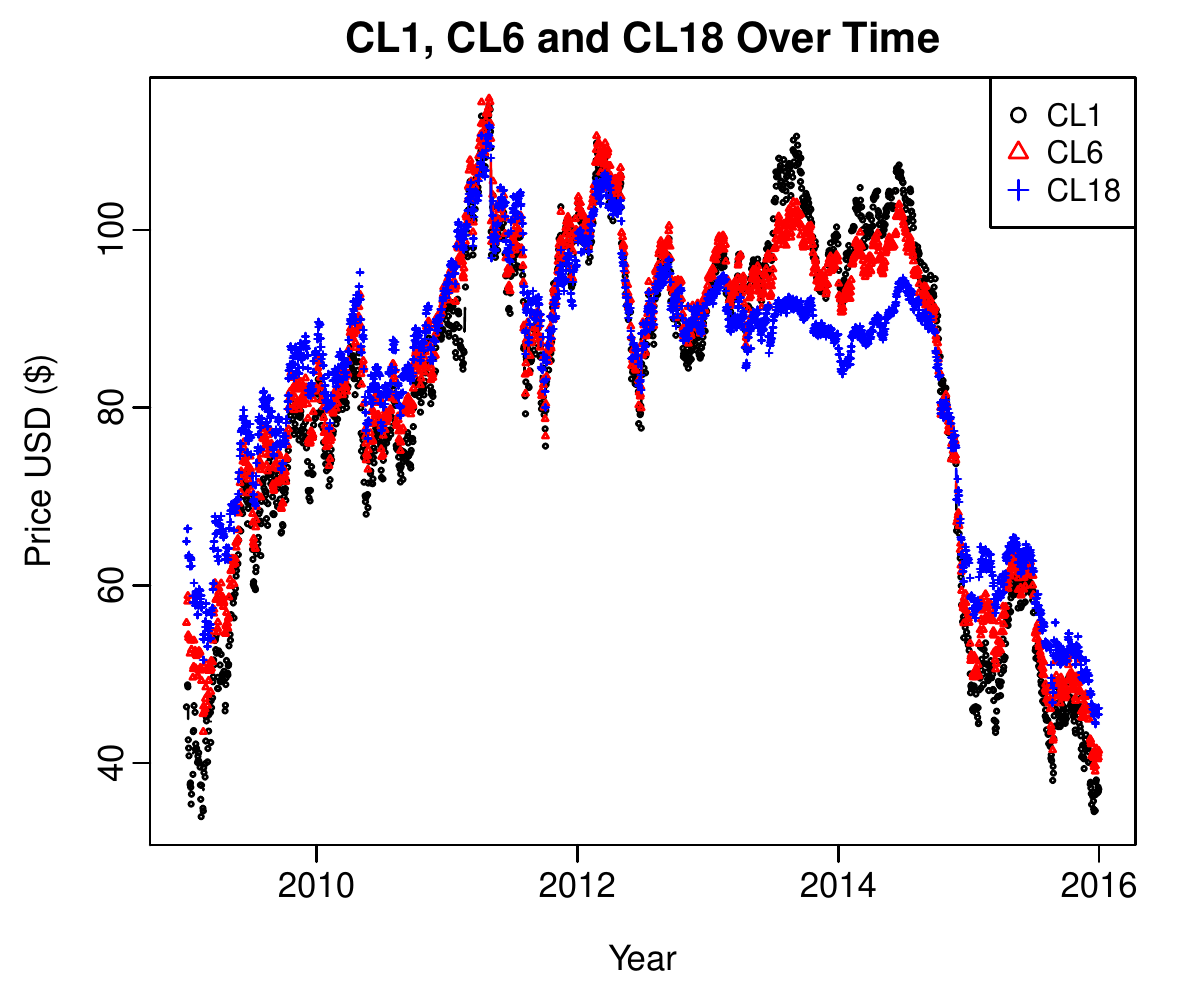}
\caption{\small Evolution of three crude oil continuous futures prices over January 2009--December 2015 period; short-term, medium-term and long-term expiries are represented by CL1, CL6 and CL18, respectively}\label{fig:2}
\end{figure}

\section{Methodology}\label{sec:3}

\subsection{Exponential Smoothing Functional Time Series Forecasts}

We adopt an FTS approach to first model and subsequently forecast the relationship between the prices associated with oil futures contracts of different expiries. To model this futures curve, we first let $(\X_t, t\in \mathbb{Z})$ be an arbitrary FTS defined on a common probability space $(\Omega, A, P)$. In our case $\X_t$ represents the futures curve on each day $i$, with the continuum $\theta$ given in terms of the futures contract expiry, CL1, CL2, etc.\footnote{In this paper we use the generic rolling contracts, CL1, CL2, etc., for consistency with prior studies including those that propose our adopted benchmark models. However, the FTS framework we propose could also be modified to handle each futures contract's precise number of days to expiry.} Each function is a square integrable function satisfying $\|\X_t\|^2 = \int_{\mathcal{I}}\X_t^2(\theta)d\theta<\infty$.\footnote{It is assumed that the observations $\X_t$ are elements of the Hilbert space $\mathcal{H} = \mathcal{L}^2(\mathcal{I})$ equipped with the inner product $\langle w, v\rangle = \int_{\mathcal{I}}w(\theta)v(\theta)d\theta$, where $\theta$ represents a continuum and $\mathcal{I}\subset R$ denotes a function support range and $R$ denotes the real line.} The notation $\X\in \mathcal{L}_{\mathcal{H}}^p(\Omega, A, P)$ is used to indicate that $\text{E}(\|\X\|^p)<\infty$ for some $p>0$. For instance, when $p=1$, $\X(\theta)$ has the mean curve $\mu(\theta) = \text{E}[\X(\theta)]$; when $p=2$, the non-negative definite covariance function is given by
\begin{equation}
c_{\X}(\theta,\omega) = \text{Cov}[\X(\theta), \X(\omega)] = \text{E}\left\{[\X(\theta) - \mu(\theta)][\X(\omega) - \mu(\omega)]\right\}, \label{eq:1}
\end{equation}
for all $\theta, \omega \in \mathcal{I}$. The covariance function in~\eqref{eq:1} allows the covariance operator of $\X$ to be defined as:
\begin{equation*}
\mathcal{K}_{\X}(\phi)(\omega) = \int_{\mathcal{I}}c_{\X}(\theta,\omega)\phi(\theta)d\theta.
\end{equation*}
There then exists an orthonormal sequence $(\phi_k)$ of continuous functions and a non-increasing sequence $(\lambda_k)$ of positive numbers, such that:
\begin{equation*}
c_{\X}(\theta,\omega) = \sum^{\infty}_{k=1}\lambda_k\phi_k(\theta)\phi_k(\omega), \qquad \theta, \omega \in \mathcal{I}.
\end{equation*}
Therefore, a Karhunen-Lo\`{e}ve expansion of such a stochastic process $\X(\theta)$ can be expressed as:
\begin{align}
\X(\theta) &= \mu(\theta) + \sum^{\infty}_{k=1}\beta_k\phi_k(\theta) \notag\\
&= \mu(\theta) + \sum^K_{k=1}\beta_k\phi_k(\theta) + e(\theta), \label{eq:2}
\end{align}
where scores $\beta_k$ are given by the projection of $\X(\theta) - \mu(\theta)$ in the direction of the $k^{\text{th}}$ eigenfunction $\phi_k$; that is, $\beta_k = \langle \X(\theta) - \mu(\theta), \phi_k(\theta)\rangle$; $e(\theta)$ denotes the model truncation error function with a mean of zero and a finite variance; and $K<n$ is the number of retained components. Equation~\eqref{eq:2} facilitates dimension reduction as the first $K$ terms often provide a good approximation to the infinite sums, and thus the information inherent in $\bm{\X}(\theta)$ can be adequately summarised by the $K$-dimensional vector $(\bm{\beta}_1,\cdots,\bm{\beta}_K)$.

There are at least four common approaches to determine our selection of $K$, namely cross-validation \citep{RS05}, Akaike's information criterion \citep{Akaike74}, the bootstrap method \citep{HV06}, the ratio method \citep{LY12} and explained variance \citep{Chiou12}. Here, the value of $K$ is determined as the minimum that reaches a certain level of the proportion of total variance explained by the $K$ leading components such that:
\begin{equation*}
K = \argmin_{K: K\geq 1}\left\{\sum^K_{k=1}\widehat{\lambda}_k\Big/\sum^{\infty}_{k=1}\widehat{\lambda}_k\mathds{1}_{\left\{\widehat{\lambda}_k>0\right\}}\geq P_1\right\},
\end{equation*}
where $P_1 = 99\%$, $\mathds{1}_{\left\{\widehat{\lambda}_k>0\right\}}$ is to exclude possible zero eigenvalues, and $\mathds{1}\{\cdot\}$ represents the binary indicator function. When we align with \cite{CS08} and \cite{CDK16} and utilise three principal components, $99\%$ of the variation in our dataset is explained; therefore we also specify $P_1 = 99\%$.

By characterising the relationship between oil futures prices of different expiries in the above manner, we obtain daily functional realisations of its term structure, or futures curve. With a set of realisations $\bm{\X}(\theta) = \{\X_1(\theta),\dots,\X_n(\theta)\}$, the mean function can be approximated as $\widehat{\mu}(\theta) = \sum^n_{t=1}\X_t(\theta)/n$, and the theoretical covariance function can be empirically estimated as:
\begin{equation*}
\widehat{c}_n = \frac{1}{n}\sum^n_{i=1}\langle \X_t, x\rangle \X_t, \qquad x\in \mathcal{H},
\end{equation*}
which is shown to be consistent under weak dependency \citep[see, e.g.,][]{HK10}. From the empirical covariance function, we can extract estimated functional principal component functions $\bm{B} = \{\widehat{\phi}_1(\theta),\dots,\widehat{\phi}_K(\theta)\}$. Conditioning on the past functions $\bm{\X}(\theta)$ and the estimated functional principal components $\bm{B}$, the one-step-ahead point forecast of $\X_{n+1}(\theta)$ can be expressed as:
\begin{align*}
\widehat{\X}_{n+1|n}(\theta) &= \text{E}[\X_{n+1}(\theta)|\bm{\X}(\theta), \bm{B}] \\
&= \widehat{\mu}(\theta) + \sum^K_{k=1}\widehat{\beta}_{n+1|n,k}\widehat{\phi}_k(\theta),
\end{align*}
where $\widehat{\beta}_{n+1|n,k}$ denotes time series forecasts of the $k^{\text{th}}$ component scores. The forecasts of these scores can be obtained via a univariate time series forecasting method, such as exponential smoothing \citep[see][for a review]{OF13}, which can handle non-stationarity of the functional principal component scores.

Following early work by \cite{GM85}, we consider a damped trend exponential smoothing model, where a parameter $0\leq \xi<1$ dampens the trend to a flat line sometime in the future. For each set of principal component scores, the resulting model is expressed as:
\begin{align*}
\beta_t &= \ell_{t-1} + \xi b_{t-1} + \epsilon_t, \\
\ell_t &= \ell_{t-1} + \xi b_{t-1} + \delta \epsilon_t,  \\
b_t &= \xi b_{t-1} + \gamma\epsilon_t,
\end{align*}
where $\ell_t$ denotes level at time $t=1,\dots,n$, $b_t$ denotes the growth rate and $\epsilon_t$ denotes the error term. \cite{EGS06} also consider exponential smoothing for financial economic forecasting.

\subsection{Benchmarks}

To evaluate the performance of our proposed FTS framework, we employ the use of fundamental and latent factor benchmarks. We utilise fundamental factors proposed by \cite{ABNU16} to model commodity futures prices, namely the Trade Weighted US Dollar Exchange Rate, S\&P 500 Index levels, the VIX (a measure of the market's expectation of future S\&P 500 index volatility) and the EcPol developed by \cite{BBD15}. The model is specified as follows:
\begin{equation*}
\text{Oil}_t^{\tau}=\zeta+\beta_{1}\text{SP500}_{t-1}+\beta_{2}\text{VIX}_{t-1}+\beta_{3}\text{USD}_{t-1}+\beta_{4}\text{EcPol}_{t-1}+\varepsilon_{t}, 
\end{equation*}
with $\text{Oil}^{\tau}_t$ denoting the log return of the oil futures with expiry $\tau$ at time $t$, $\zeta$ denoting the intercept term, SP500$_{t-1}$ denoting S\&P 500 log return at time $t-1$, VIX$_{t-1}$ denoting VIX volatility index log change at time $t-1$, USD denoting the trade-weighted US dollar index at time $t-1$, EcPol$_{t-1}$ denoting US Economic Policy Uncertainty Index log change at time $t-1$ and $\varepsilon_t$ denoting the error term at time $t$. We hereafter refer to this model as Fundamental.

Our second benchmark is the latent factor model, as proposed by \cite{CS08}. They use three discrete principal component extractions as factors for forecasting, as follows:
\begin{equation*}
\text{Oil}_{t}^{\tau}=\zeta+\beta_{1}\text{f}_{1, t-1}+\beta_{2}\text{f}_{2, t-1}+\beta_{3}\text{f}_{3, t-1}+\upsilon_{t}, 
\end{equation*}
where $\text{f}_{1,t-1}, \text{f}_{2,t-1}$ and $\text{f}_{3,t-1}$ denote the first, second and third principal components calculated from the oil futures prices, $\beta_1, \beta_2$ and $\beta_3$ denote the coefficients and $\upsilon_t$ denotes the error term. We hereafter refer to this model as PC.

Finally, as a na\"{i}ve benchmark, we also incorporate the random walk without drift into our in-sample analysis. The random walk without drift model can be expressed as:
\begin{equation*}
\text{Oil}_{n+1}^{\tau} = \text{Oil}_{n}^{\tau} + \nu_{n+1},
\end{equation*}
where $\nu_{n+1}$ is a zero-mean error term that is uncorrelated to $\text{Oil}_n^{\tau}$. The random walk without drift is also utilised in the out-of-sample performance evaluation to produce the Mean Absolute Scaled Error (MASE) as proposed by \cite{HK06}. The MASE loss function is further detailed in the following sections.

\subsection{Forecast Evaluation}

Out-of-sample forecasts are obtained using an expanding scheme. Each day an additional observation is added to an expanding training window with the models being re-estimated. We choose to expand the training set and re-estimate the model at each time step, daily, to incorporate all available up-to-date information into our forecast.\footnote{Alternatively, a rolling window estimation approach that would iteratively discard earlier data points could be adopted. However, we proceed on the basis that all data points are relevant in the estimation stage with the exponential smoothing forecasts weighting the most recent data points more heavily when producing the forecasts.} Using our January 2009--December 2015 dataset, we statistically compare the model forecasts using a 500-day out-of-sample window, whereby the January 2009--December 2013 period is initially designated as the training window with the calculated forecasts being compared with the 500 unknown observations in the December 2013--December 2015 window. As an additional robustness check that controls for sensitivity to the specific out-of-sample period, 250-day and 750-day out-of-sample windows are also specified. We evaluate in-sample predictions using (where appropriate) a variety of performance measures: Mean Absolute Error (MAE), Mean Mixed Error Under (MME(U)), Mean Mixed Error Over (MME(O)), and Mean Correct Predictor of the Direction of Change (MCPDC).

\subsection{Loss Functions}

We assess the performance of the models using the following measures:
\begin{enumerate}
\item[1)] MAE is the average of the absolute differences between the prediction, ${\textstyle \widehat{\text{Oil}}_{n+\vartheta|n+\vartheta-1}}$, and the corresponding observation, ${\text{Oil}_{n+\vartheta}}$. It measures the average error magnitude of the predictions, regardless of error direction and serves to aggregate errors into a single measure of predictive power:
\begin{align*}
\text{MAE}^{\tau}&=\frac{1}{T}\sum_{\vartheta=1}^T \left|\text{Oil}^{\tau}_{n+\vartheta}-\textstyle \widehat{\text{Oil}}^{\tau}_{n+\vartheta|n+\vartheta-1}\right|,\qquad \tau = 1,\dots,11, \\
\text{MAE}_{\text{overall}} &= \frac{1}{11}\sum^{11}_{\tau=1}\text{MAE}^{\tau},
\end{align*}
where ${\text{Oil}^{\tau}_{n+\vartheta}}$ and ${\textstyle \widehat{\text{Oil}}^{\tau}_{n+\vartheta|n+\vartheta-1}}$ are the observed values and their predictions for a particular expiry and model, and $T$ denotes the number of observations in the forecasting period.
\item[2)] Mean error (ME) is the average of the differences between the prediction, ${\textstyle \widehat{\text{Oil}}_{n+\vartheta|n+\vartheta-1}}$, and the corresponding observation, ${\text{Oil}_{n+\vartheta}}$. It measures the average error magnitude of the predictions, taking into account error direction, and serves to aggregate errors into a single measure of predictive power:
\begin{align*}
\text{ME}^{\tau}&=\frac{1}{T}\sum_{\vartheta=1}^T \left(\text{Oil}^{\tau}_{n+\vartheta}-\textstyle \widehat{\text{Oil}}^{\tau}_{n+\vartheta|n+\vartheta-1}\right),\qquad \tau = 1,\dots,11, \\
\text{ME}_{\text{overall}} &= \frac{1}{11}\sum^{11}_{\tau=1}\text{ME}^{\tau},
\end{align*}
where ${\text{Oil}^{\tau}_{n+\vartheta}}$ and ${\textstyle \widehat{\text{Oil}}^{\tau}_{n+\vartheta|n+\vartheta-1}}$ are the observed values and their predictions for a particular expiry and model, and $T$ denotes the number of observations in the forecasting period.
\item[3)] MASE utilises the error produced by an in-sample na\"{i}ve random walk without drift to scale and compare the out-of-sample errors observed. It measures the error magnitude scaled by the magnitude of the in-sample error:
\begin{equation*}
q_{\vartheta}^{\tau} = \frac{\text{Oil}_{n+\vartheta}^{\tau} - \widehat{\text{Oil}}^{\tau}_{n+\vartheta|n+\vartheta-1}}{\frac{1}{n-1}\sum^n_{\varphi=2}|\text{Oil}_{\varphi} - \text{Oil}_{\varphi-1}|}, \qquad \vartheta = 1,\dots, T.
\end{equation*}
The error metric is independent of the scale of the data. A scaled error is less than one if it produces a better forecast than the average one-step, na\"{i}ve forecast computed in-sample \citep{HK06}. The MASE is simply:
\begin{equation*}
\text{MASE} = \frac{1}{T\tau}\sum^{T}_{\vartheta=1}\sum^{11}_{\tau=1}\left|q_{\vartheta}^{\tau}\right|.
\end{equation*}
\item[4)] MME is an asymmetric loss function. MME(U) penalises under-predictions more heavily, while MME(O) penalises over-predictions more heavily. It is not a scoring rule in the conventional sense. However, it is useful in that it provides us with an indication of the tendency of a model to under- or over-predict. This is an important consideration for consumers of oil, as under-predictions of future oil prices may be of greater concern to those in need of future supply:
\begin{align*}
\text{MME}^{\tau}\text{(U)} &= \frac{1}{T}\left[\sum_{\psi = \eta_1^{\text{O}}}^{\eta_m^{\text{O}}}\left|\text{Oil}^{\tau}_{n+\psi}-\textstyle \widehat{\text{Oil}}^{\tau}_{n+\psi|n+\psi-1}\right|+\sum_{\psi = \eta_1^{\text{U}}}^{\eta_m^{\text{U}}}\sqrt{\left|\text{Oil}^{\tau}_{n+\psi}-\textstyle \widehat{\text{Oil}}^{\tau}_{n+\psi|n+\psi-1}\right|}\right], \\
\text{MME}^{\tau}\text{(O)} &= \frac{1}{T}\left[\sum_{\psi = \eta_1^{\text{U}}}^{\eta_m^{\text{U}}}\left|\text{Oil}^{\tau}_{n+\psi}-\textstyle \widehat{\text{Oil}}^{\tau}_{n+\psi|n+\psi-1}\right|+\sum_{\psi = \eta_1^{\text{O}}}^{\eta_m^{\text{O}}}\sqrt{\left|\text{Oil}^{\tau}_{n+\psi}-\textstyle \widehat{\text{Oil}}^{\tau}_{n+\psi|n+\psi-1}\right|}\right], \\
\text{MME}(\text{U})_{\text{overall}} &= \frac{1}{11}\sum^{11}_{\tau=1}\text{MME}^{\tau}\text{(U)}, \\
\text{MME}(\text{O})_{\text{overall}} &= \frac{1}{11}\sum^{11}_{\tau=1}\text{MME}^{\tau}\text{(O)},
\end{align*}
where $\eta_m^{\text{U}}$ denotes the number of under-predictions and $\eta_m^{\text{O}}$ denotes the number of over-predictions. $\big\{\eta_1^{\text{O}},\dots,\eta_m^{\text{O}}\big\}$ represents the indices of the over-predictions, and $\big\{\eta_1^{\text{U}},\dots,\eta_m^{\text{U}}\big\}$ represents the indices of the under-predictions. 
\item[5)] The MCPDC is the percentage of predictions for which the prediction, ${\textstyle \widehat{\text{Oil}}_{n+\vartheta|n+\vartheta-1}}$, has the same sign as the corresponding observation, ${\text{Oil}_{n+\vartheta}}$. MCPDC measures how well the model can predict the direction of movement, regardless of error magnitude. 
\end{enumerate}

When evaluating out-of-sample forecasts, as opposed to in-sample predictions, we refer to the above measures as Mean Absolute Forecast Error (MAFE), Mean Forecast Error (MFE), Mean Absolute Scaled Forecast Error (MASFE), Mean Mixed Forecast Error Under (MMFE(U)), Mean Mixed Forecast Error Over (MMFE(O)), and Mean Correct Forecast of the Direction of Change (MCFDC), respectively.

\section{Empirical Results}\label{sec:4}

\subsection{In-sample modelling}

\begin{figure}[!tbp]
\centering
\includegraphics[width=17.5cm]{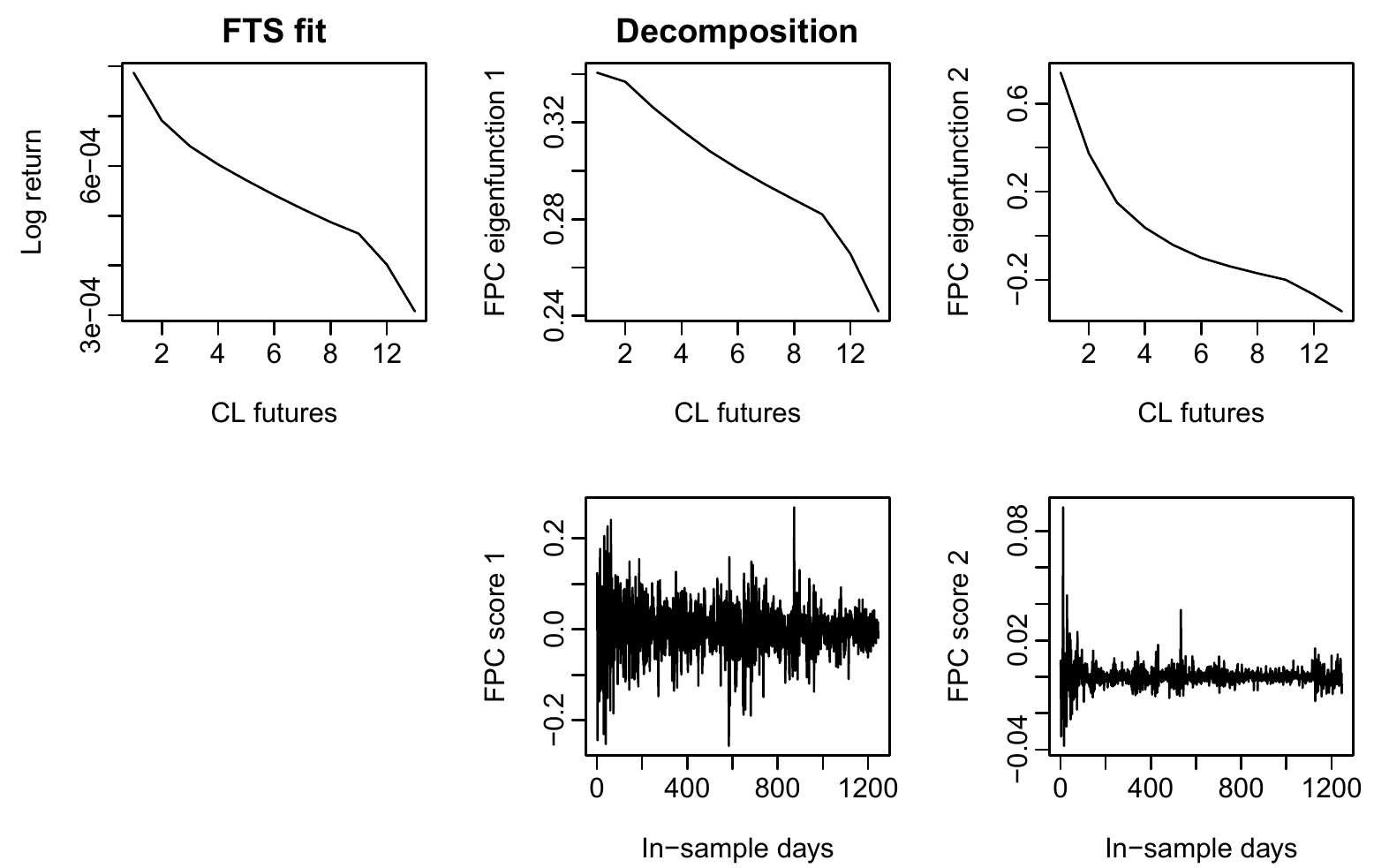}
\caption{\small FTS in-sample fit: the mean fitted FTS (Functional Time Series) model and a graphical representation of its FPC (Functional Principal Component) decomposition over the January 2009--December 2013 period}\label{fig:3}
\end{figure}

We briefly outline in-sample fit before drawing inferences using measures of out-of-sample performance. First, we outline the number of principal components used. In line with \cite{CS08} and \cite{CDK16} we specify three discrete principal components to model the oil futures curve. This results in 99\% of the variance in the data being explained. Therefore, in the functional context, we also seek to explain 99\% of the total variation, resulting in two functional principal components being adapted for the FTS model. Interestingly, a third functional principal component contributes less than 0.1\% explained variation. We model the relationship between the curves using the FTS decomposition shown in Figure~\ref{fig:3}. The first functional principal component can be interpreted as characterising the linear relationship between the contracts, with the second component addressing the need to model the curvature dynamic. Regression results from fitting the benchmark fundamental and PC models are available upon request.

We now seek to establish how well the models fit in an in-sample modelling exercise. In-sample modelling results using MAE, MCPDC, MME(U) and MME(O) loss functions are presented in Table~\ref{tab:2}. Each row shows the calculated performance measure on an individual expiry basis with the Overall row presenting the loss function aggregated across all contracts. In a cross-model comparison of the absolute and asymmetric performance measures the results of the in-sample fit using our training set are broadly similar. More specifically we observe the same Overall MAE result of 0.0129 across the PC, FTS and Fundamental models.\footnote{As shown in Table~\ref{tab:2} Overall in-sample MME(U)/(O) levels are also similar at 0.0562/0.0583, 0.0569/0.0575 and 0.0562/0.0582, for the PC, FTS and Fundamental models, respectively.} This indicates that despite being constructed using distinct approaches each model closely calibrates to the in-sample patterns present in the dataset. Furthermore, this is in line with \cite{CDK16} who find no discernible in-sample difference between the competing oil futures models. Interestingly, all models considered outperform the na\"{i}ve random walk which led to an Overall MAE of 0.0188. Another notable result is that while all models correctly predict the direction of change better than would be achieved by chance alone (i.e., MCPDC$>0.5$), the FTS boasts the greatest overall success, at 53.38\%, versus 52.70\% for the Fundamental and 52.95\% for the PC models.

\begin{table}[!htbp]
\tabcolsep 0.21in
\begin{center}
\caption{\small In-sample Loss Functions\label{tab:2}}
\begin{tabular}{@{}lcccccccc@{}}\toprule
Expiry & \multicolumn{4}{c}{MAE} & & \multicolumn{3}{c}{MCPDC} \\
& PC & FTS & Fund & RW & & PC & FTS & Fund \\\midrule
Overall & 0.0129 & 0.0129 & 0.0129 & 0.0188 && 0.5295 & 0.5338 & 0.5270      \\
CL1     & 0.0152 & 0.0152 & 0.0152 & 0.0220     && 0.5325 & 0.5245 & 0.5277      \\
CL2     & 0.0144 & 0.0145 & 0.0145 & 0.0209     && 0.5398 & 0.5213 & 0.5229      \\
CL3     & 0.0139 & 0.0139 & 0.0139 & 0.0202     && 0.5261 & 0.5277 & 0.5205      \\
CL4     & 0.0135 & 0.0135 & 0.0135  & 0.0197    && 0.5277 & 0.5285 & 0.5213      \\
CL5     & 0.0131 & 0.0132 & 0.0131  & 0.0192    && 0.5301 & 0.5333 & 0.5237      \\
CL6     & 0.0128 & 0.0128 & 0.0128 & 0.0187     && 0.5333 & 0.5357 & 0.5349      \\
CL7     & 0.0125 & 0.0125 & 0.0125 & 0.0183     && 0.5285 & 0.5373 & 0.5333      \\
CL8     & 0.0123 & 0.0123 & 0.0123 & 0.0180     && 0.5261 & 0.5365 & 0.5309      \\
CL9     & 0.0120 & 0.0120 & 0.0120 & 0.0176      && 0.5253 & 0.5414 & 0.5317      \\
CL12    & 0.0114 & 0.0114 & 0.0114 & 0.0167      && 0.5245 & 0.5414 & 0.5269      \\
CL18    & 0.0105 & 0.0105 & 0.0105 & 0.0154     && 0.5309 & 0.5446 & 0.5229      \\
\midrule
Expiry & \multicolumn{3}{c}{MME(U)}  && & \multicolumn{3}{c}{MME(O)} \\
& PC & FTS & Fund && & PC & FTS & Fund \\\midrule
Overall & 0.0562 & 0.0569 & 0.0562 &&& 0.0583 & 0.0575 & 0.0582 \\
CL1     & 0.0628 & 0.0637 & 0.0626 &&& 0.0627 & 0.0618 & 0.0627 \\
CL2     & 0.0608 & 0.0617 & 0.0609 &&& 0.0615 & 0.0606 & 0.0615 \\
CL3     & 0.0592 & 0.0600 & 0.0594 &&& 0.0605 & 0.0597 & 0.0605 \\
CL4     & 0.0581 & 0.0588 & 0.0583 &&& 0.0596 & 0.0589 & 0.0596 \\
CL5     & 0.0570 & 0.0577 & 0.0570 &&& 0.0590 & 0.0583 & 0.0589 \\
CL6     & 0.0560 & 0.0566 & 0.0560 &&& 0.0583 & 0.0576 & 0.0582 \\
CL7     & 0.0550 & 0.0558 & 0.0551 &&& 0.0578 & 0.0571 & 0.0576 \\
CL8     & 0.0543 & 0.0550 & 0.0543 &&& 0.0573 & 0.0566 & 0.0571 \\
CL9     & 0.0536 & 0.0543 & 0.0536 &&& 0.0568 & 0.0561 & 0.0566 \\
CL12    & 0.0519 & 0.0525 & 0.0519 &&& 0.0552 & 0.0545 & 0.0550 \\
CL18    & 0.0493 & 0.0498 & 0.0495 &&& 0.0526 & 0.0519 & 0.0523 \\
\bottomrule
\end{tabular}
\end{center}
{\small \textbf{Note:} For each of the 11 different expiry continuous crude oil futures in our sample, this table presents the calculated performance measures (MAE, MME(U/O), and MCPDC) for the FTS, Fund (Fundamental) and PC models fitted to our daily dataset during the January 2009--December 2013 in-sample period. Overall refers to the loss function aggregated across all considered expiries. RW represents the random walk without drift. Because of its construction, it is not appropriate to calculate MCPDC and MME(U/O) for RW.}
\end{table}

\subsection{Out-of-sample forecasts}

To overcome the potential for in-sample data overfitting we derive inferences from our out-of-sample testing results. The out-of-sample framework aims to answer the question; how well do the models perform in terms of producing accurate forecasts of unknown future observations? We choose to employ a recursively expanding estimation window as it closely represents a situation a practitioner might find themselves in, whereby they use all available data up to today to forecast tomorrow's price. An alternative approach is to use a rolling fixed-length estimation window. We argue, however, that our expanding window approach represents a more appropriate environment for the exponential smoothing framework to determine its weightings, given that it internally deems observations in the distant past to have less of an effect on current forecasts.

Table~\ref{tab:3} presents the loss functions produced under our out-of-sample exercise. To produce the dynamic iterative forecasts, the \verb ftsa \ package of \cite{HS17} was employed. The FTS model shows a lower forecasted absolute error overall, with its forecasting improvement not being confined to any one segment of the futures curve. The MFE loss function also indicates that the FTS has the lowest error, both overall--at -0.0010 versus -0.0017 and -0.0016, for FTS versus PC and Fundamental, respectively--and across each of the individual contract maturities. Furthermore, the use of the MASFE measure results in a similar conclusion with an error measure of 0.7286 overall for FTS versus 0.7324 and 0.7382, for PC and Fundamental, respectively. When we focus on a purely directional loss function, the FTS model correctly predicts the direction of price change more often than PC and Fundamental, and, very importantly, more often than by chance alone (i.e., MCFDC$>0.5$) for each expiry. Analysing the asymmetric MMFE(U) and MMFE(O) loss functions shows us that the FTS model has a comparative tendency to overestimate the price change--information that might be useful for instance, when adopting such forecasting frameworks for use in hedging or trading applications.\footnote{As shown in Table~\ref{tab:3}, Overall out-of-sample MMFE(U)/(O) stand at 0.0646/0.0530, 0.0624/0.0549 and 0.0647/0.0536, for PC, FTS and Fundamental, respectively.} The FTS model's comparative tendency to overestimate price changes might result in contrasting implications for those adopting it for hedging or speculative trading activities.\footnote{First, consider an airline as an example of a potential commercial hedger. With oil being an incurred operational cost, the use of a model that overestimates projected oil price changes, and hence increased volatility of future costs, might lead to management being too cautious in their business decisions. Second, consider an oil option trader as a potential speculative trader. If they were to adopt a model that overestimates underlying oil futures price changes, it could lead to them displaying too little caution and overpaying for the associated option. This is caused by price volatility in the underlying futures contract being a key option pricing component.} Finally, using 250 and 750 days as alternative out-of-sample window lengths leads to qualitatively similar results being obtained.

\tabcolsep 0.265in
\begin{longtable}{@{}lccccccc@{}}
\caption{\small Out-of-sample Loss Functions} \label{tab:3} \\
\toprule
Expiry & \multicolumn{3}{c}{MAFE} & & \multicolumn{3}{c}{MCFDC} \\
& PC & FTS & Fund & & PC & FTS & Fund \\
\midrule
\endfirsthead
\toprule
Expiry & \multicolumn{3}{c}{MAFE} & & \multicolumn{3}{c}{MCFDC} \\
& PC & FTS & Fund & & PC & FTS & Fund \\
\midrule
\endhead

\hline \multicolumn{8}{r}{{Continued on next page}} \\ 
\endfoot
\endlastfoot
Overall	& 0.0138 & 	\textBF{0.0137} & 	0.0139 & & 		0.5087	& \textBF{0.5169}	& 0.4831 \\
CL1	& \textBF{0.0166}	& \textBF{0.0166} & 	0.0168 & 	& 	\textBF{0.5480}& 	0.5060	& 0.4640 \\
CL2	& \textBF{0.0160}	& \textBF{0.0160} & 	0.0162& 	& 	\textBF{0.5440}& 	0.5160	& 0.4640 \\
CL3	& 0.0154	& \textBF{0.0153}	& 0.0156	& &	0.5020	& \textBF{0.5180}& 	0.4660 \\
CL4	& 0.0149	& \textBF{0.0148}	& 0.0151	& &	0.4960	& \textBF{0.5220}& 	0.4820 \\
CL5	& 0.0144	& \textBF{0.0143}	& 0.0145	& &	0.4960& 	\textBF{0.5220}& 	0.4820 \\
CL6	& 0.0138	& \textBF{0.0137}& 	0.0139	& &	0.4920	& \textBF{0.5160}	& 0.4860 \\
CL7	& 0.0135& 	\textBF{0.0134}	&0.0135	&&	0.4960	&\textBF{0.5280}	&0.4960 \\
CL8	& 0.0131	& \textBF{0.0130}	& 0.0131 &	&	0.5080 &	\textBF{0.5220} &	0.5020 \\
CL9	& 0.0126	& \textBF{0.0125}	& 0.0126	&&	0.5020	& \textBF{0.5220}	& 0.4880 \\
CL12 & \textBF{0.0115} &	\textBF{0.0115} &	0.0116	&&	0.4960	& \textBF{0.5100} &	0.4920 \\
CL18 & 0.0097	& \textBF{0.0096}	& 0.0097	& &	\textBF{0.5160}	& 0.5040	& 0.4920 \\
\midrule
Expiry & \multicolumn{3}{c}{MASFE} & & \multicolumn{3}{c}{MFE} \\
& PC & FTS & Fund & & PC & FTS & Fund \\
\midrule
Overall & 0.7324 & \textBF{0.7286} & 0.7382      &  & -0.0017 & \textBF{-0.0010} & -0.0016     \\
CL1     & \textBF{0.7548} & 0.7551 & 0.7656      &  & -0.0021 & \textBF{-0.0012} & -0.0021     \\
CL2     & 0.7628 & \textBF{0.7627} & 0.7737      &  & -0.0019 & \textBF{-0.0011} & -0.0019     \\
CL3     & 0.7621 & \textBF{0.7587} & 0.7706      &  & -0.0018 & \textBF{-0.0010} & -0.0018     \\
CL4     & 0.7575 & \textBF{0.7526} & 0.7640      &  & -0.0018 & \textBF{-0.0010} & -0.0017     \\
CL5     & 0.7507 & \textBF{0.7452} & 0.7557      &  & -0.0017 & \textBF{-0.0010} & -0.0017     \\
CL6     & 0.7389 & \textBF{0.7335} & 0.7424      &  & -0.0017 & \textBF{-0.0010} & -0.0016     \\
CL7     & 0.7356 & \textBF{0.7290} & 0.7386      &  & -0.0016 & \textBF{-0.0009} & -0.0016     \\
CL8     & 0.7267 & \textBF{0.7220} & 0.7304      &  & -0.0016 & \textBF{-0.0009} & -0.0015     \\
CL9     & 0.7133 & \textBF{0.7077} & 0.7164      &  & -0.0015 & \textBF{-0.0009} & -0.0015     \\
CL12    & 0.6918 & \textBF{0.6873} & 0.6939      &  & -0.0014 & \textBF{-0.0008} & -0.0014     \\
CL18    & 0.6261 & \textBF{0.6228} & 0.6297      &  & -0.0013 & \textBF{-0.0007} & -0.0012  \\ 
\midrule
Expiry & \multicolumn{3}{c}{MMFE(U)} & & \multicolumn{3}{c}{MMFE(O)} \\
& PC & FTS & Fund & & PC & FTS & Fund \\
\midrule
Overall &	0.0646	& \textBF{0.0624} &	0.0647 & &		\textBF{0.0530} &	0.0549	& 0.0536 \\
CL1 &	0.0728	& \textBF{0.0704} &	0.0733 & &		\textBF{0.0583} &	0.0608	& 0.0595 \\
CL2	& 0.0710 &	\textBF{0.0684} &	0.0711 &	&	\textBF{0.0574} &	0.0598 &	0.0585 \\
CL3	& 0.0694	& \textBF{0.0671} &	0.0698 & &		\textBF{0.0566} &	0.0584 & 	0.0573 \\
CL4	& 0.0682 &	\textBF{0.0659} &	0.0684 & &		\textBF{0.0558}	& 0.0574 &	0.0563 \\
CL5	& 0.0667	& \textBF{0.0644}	& 0.0668	& &	\textBF{0.0545} &	0.0561 &	0.0549 \\
CL6	& 0.0649 &	\textBF{0.0627} &	0.0648	& &	\textBF{0.0537} &	0.0555 &	0.0542 \\
CL7	& 0.0643	& \textBF{0.0619}	& 0.0641	& &	\textBF{0.0527} &	0.0541 &	0.0530 \\
CL8	& 0.0628 &	\textBF{0.0608} &	0.0627	& &	\textBF{0.0517} &	0.0533	 & 0.0520 \\
CL9	& 0.0610 &	\textBF{0.0591} &	0.0612	& &	\textBF{0.0509} &	0.0523	& 0.0511 \\
CL12 &	0.0579 &	\textBF{0.0560} &	0.0578 & &		\textBF{0.0484} &	0.0502 &	0.0488 \\
CL18 &	0.0521	& \textBF{0.0499} &	0.0518 & &		\textBF{0.0436} &	0.0458 & 	0.0443 \\\bottomrule
\end{longtable}
\noindent {\small \textbf{Note:} For each of the 11 different expiry continuous crude oil futures in our sample, this table presents the calculated forecast performance measures (MAFE, MASFE, MFE, MMFE(U/O) and MCFDC) for the out-of-sample forecasts produced by our FTS, Fund (Fundamental) and PC frameworks during the December 2013--December 2015 period. Overall refers to the loss function aggregated across all 11 expiries.  The lowest forecast errors are shown in bold.}
\\

As these measures are simply indicators of an improvement in performance, we need to establish if the FTS model is indeed statistically superior. To this end, we turn to the MCS results presented in Table~\ref{tab:4}, where model confidence sets are constructed at confidence levels of 75\% and 90\% in line with \cite{HLN11}. Details of the MCS framework are presented in the Appendix. As also outlined there, absolute error is adapted to distinguish between models in terms of out-of-sample performance. From the Overall row, the FTS resides exclusively in the superior set of models across the full futures curve, and for both test statistics. Furthermore, when focusing on each of the individual futures expiries, the FTS is classified, although not always exclusively, as a superior model in each case. Our MCS approach robustly identifies model outperformance, with its design inherently controlling for multiple comparisons. Multiple comparisons is the issue that given a sufficient number of simultaneous tests seemingly significant outperformance will be uncovered by chance alone.

\begin{table}[!htbp]
\begin{center}
\tabcolsep 0.133in
\caption{\small Model Confidence Set Results\label{tab:4}}
\begin{tabular}{@{}llcccclccc@{}}\toprule
& \multicolumn{4}{c}{$\alpha=0.25$} & & \multicolumn{4}{c}{$\alpha=0.1$} \\\cmidrule{3-5}\cmidrule{8-10}
Statistic & Expiry	& PC    	 	& FTS       	& Fundamental	 & & Expiry	& PC    	 	& FTS       	& Fundamental	 	 \\ \midrule
$T_{\text{R,M}}$ & Overall & & $\dagger$ & & & Overall & &  $\ddagger$ & \\
& CL1 & $\dagger$ & $\dagger$ & & & CL1 & $\ddagger$ & $\ddagger$ & \\
& CL2 & $\dagger$ & $\dagger$ & & & CL2 & $\ddagger$ & $\ddagger$ & \\
& CL3 & $\dagger$ & $\dagger$ & & & CL3 & $\ddagger$ & $\ddagger$ & \\ 
& CL4 & 		      & $\dagger$ & & & CL4 & $\ddagger$ & $\ddagger$ & \\  
& CL5 & 		      & $\dagger$ & & & CL5 & $\ddagger$ & $\ddagger$ & \\  
& CL6 & 		      & $\dagger$ & & & CL6 & $\ddagger$ & $\ddagger$ & \\  
& CL7 & 		      & $\dagger$ & & & CL7 &                  & $\ddagger$ & \\  
& CL8 & 		      & $\dagger$ & & & CL8 & $\ddagger$ & $\ddagger$ & \\  
& CL9 & 		      & $\dagger$ & & & CL9 & $\ddagger$ & $\ddagger$ & \\  
& CL12 & $\dagger$ & $\dagger$ & & & CL12 & $\ddagger$ & $\ddagger$ & $\ddagger$\\  
& CL18 & $\dagger$ & $\dagger$ & & & CL18 & $\ddagger$ & $\ddagger$ & \\\midrule
$T_{\text{max},\text{M}}$ & Overall 	&  	& $\dagger$         	&             		 & & Overall 	&     & $\ddagger$      	&             		 \\
& CL1     	& $\dagger$  	& $\dagger$ &             		&          			& CL1     	& $\ddagger$ 	& $\ddagger$&             		 \\
& CL2     	& $\dagger$  	& $\dagger$ &             		  &         			& CL2     	& $\ddagger$ 	& $\ddagger$&             		 \\
& CL3     	& $\dagger$ 	&  $\dagger$        	&            	              		& 	& CL3     	& $\ddagger$ 	& $\ddagger$&             		 \\
& CL4     	&  	& $\dagger$         	&            		          	& 		& CL4     	& $\ddagger$	&  $\ddagger$        	&             		 \\
& CL5     	&  	&  $\dagger$        	&             		        		& 		& CL5     	& $\ddagger$	& $\ddagger$         	&             		 \\
& CL6     	&   & $\dagger$ &      	         & 			& CL6     	& $\ddagger$ 	& $\ddagger$&     	 \\
& CL7     	&   & $\dagger$ &      	         & 			& CL6     	& $\ddagger$ 	& $\ddagger$&     	 \\
& CL8     	&  	& $\dagger$ & & 			& CL8     	& $\ddagger$	& $\ddagger$& \\
& CL9     	&  	& $\dagger$ & & 			& CL9     	& $\ddagger$ 	& $\ddagger$& \\
& CL12    	& $\dagger$  	& $\dagger$ &      	          & 			& CL12    	& $\ddagger$ 	& $\ddagger$& \\
& CL18    	& $\dagger$	& $\dagger$          	&             		 & & CL18    	& $\ddagger$ 	& $\ddagger$&             		 \\\bottomrule
\end{tabular}
\end{center}
{\small \textbf{Note:} For each of the 11 different expiry continuous crude oil futures in our sample, this table presents the results of the model confidence test of \cite{HLN11} based on the out-of-sample forecasts produced by PC, our FTS and Fundamental methods during the December 2013--December 2015 period. Overall refers to forecasts aggregated across each of the 11 expiries: $\dagger$ and $\ddagger$ are used to indicate that the framework resides in the superior set of models at 75\% and 90\% confidence levels, respectively.}
\end{table}

Table~\ref{tab:5} presents the results from alternative forecast evaluation frameworks to show the robustness of our results to the choice of test selected. All of the pairwise tests of our proposed functional approach versus the Fundamental or PC models indicate that our FTS model outperforms at the 1\% significance level. This is consistent when employing the test of predictive accuracy in \cite{DM95}, the modified version in \cite{HLN97} and the test of conditional predictive ability in \cite{GW06}. When we focus on the collective tests that simultaneously evaluate all models with respect to a reference benchmark model, we again verify that the forecast produced by the FTS model is a significant improvement over those of the other models considered. More specifically, when we specify the PC model as a benchmark, the superiority is consistent at the 1\% significance level for both \citeauthor{White00}'s\citeyearpar{{White00}} reality check and \citeauthor{Hansen05}'s\citeyearpar{Hansen05} superior predictive ability tests. However, when we specify the Fundamental model as the benchmark, only the \cite{Hansen05} test is shown to be significant. This serves to highlight the increased power of the \cite{Hansen05} test over the \citeauthor{White00}'s\citeyearpar{White00} reality check on which it is based. Overall, these results provide additional evidence to support the conclusion that the FTS forecasting approach is indeed significantly better than both the PC and Fundamental models.

\begin{table}[!htbp]
\begin{center}
\tabcolsep 0.065in
\caption{\small Alternative Statistical Tests\label{tab:5}}
\begin{tabular}{@{}lccccc@{}} \toprule
& \multicolumn{2}{c}{Pairwise Tests} & & \multicolumn{2}{c}{Collective Tests} \\\cmidrule{2-3}\cmidrule{4-6}

             Test            & FTS Vs Fund & FTS Vs PC  && Fund Benchmark & PC Benchmark \\ \midrule
Diebold-Mariano (1995)     & -0.0038*** & -0.0096*** &&               &              \\
                         & (0.0030)   & (0.0077)   & &              &              \\
Modified Diebold-Mariano (1997) & -8.8250*** & -3.0326*** &&               &              \\
                         & (0.0000)   & (0.0012)   &      &         &              \\
White (2000) Reality Check        &            &       &     & 0.0000        & 0.0081***    \\
                         &            &            && (0.6910)      & (0.0000)     \\
Hansen (2005) SPA              &            &       &     & 0.6365***     & 0.1118***    \\
                         &            &            && (0.0000)      & (0.0000)     \\
Giacomini-White (2006) CPA     & 76.8000*** & 10.4300***   &&               &              \\
                         & (0.0000)   & (0.0050)   &&               &  \\\bottomrule           
\end{tabular}
\end{center}
{\small \textbf{Note:} This table presents alternative tests for statistical outperformance across all the contracts considered; $p$-values associated with each test are given in parentheses beneath the calculated test statistics. *, **, ***, represent significance at the 10\%, 5\%, and 1\% levels, respectively. The models considered are FTS, PC and Fund (Fundamental); CPA, conditional predictive ability; SPA, superior predictive ability. Please see underlying texts indicated for further technical implementation details.}
\end{table}

\section{Concluding Remarks}\label{sec:5}

Despite a strong practitioner need to model and accurately forecast oil futures prices, recent academic research has been unable to establish a discernible difference between the performance of competing models. We contribute to this strand of literature through the innovative introduction of an infinite dimensional class of factors, based on exponential smoothing forecasts of functional PCs. Our functional approach has many advantages. First, it circumvents the need to explicitly identify factors that model the shape of the oil futures curve. Second, it takes advantage of information that traditional models overlook; despite only certain contracts being actively quoted and traded on any given day, there is an underlying continuous smooth process linking oil futures contracts of different expiries. Finally, it is flexible in the sense that it is data driven, leaving the potential for it to be extended to other applications. 

Our focus on WTI crude oil futures is motivated by the critical role of derivatives trading in Europe, as highlighted by \cite{COR17}, and the growing importance of WTI crude in a European context. We present out-of-sample results that provide empirical support for the adoption of our FTS approach when forecasting WTI oil futures, with the functional model residing in the superior set of models for all considered expiries. We now outline how our results link with previous studies.

First, our in-sample results agree with \cite{CDK16}, who highlight that an in-sample fitting indicates no discernible difference between latent and macroeconomic factors when modelling oil futures. However, our proposal to employ the FTS model in the out-of-sample context contradicts their conclusions by uncovering a statistically outperforming forecasting model. Second, the contrasting in- and out-of-sample findings suggest that instability or overfitting might be influencing forecasting evaluation. These statistical phenomena have previously been highlighted by \cite{GR09} and \cite{Clark04}, respectively. Third, our results also highlight that moving beyond the discrete approaches of \cite{CS08} and \cite{ABNU16} into a continuous FTS domain yields strong empirical benefits. As well as the importance of oil price forecasts for decision making as outlined in Section~\ref{sec:1}, our uncovered forecasting outperformance could also inform economically exploitable speculative trading or hedging strategies, with their success being dependent on an individual investor's trading costs. Future work in this area might consider alternative benchmark models and forecasting horizons, seek to construct a semi-parametric model by supplementing our FTS framework with fundamental exogenous factors, or further refine forecast accuracy through a forecast combination approach such as employed by \cite{GN18}.

\if0\blind
{
\section*{Acknowledgments}

The authors are grateful for the insightful comments and suggestions of the participants at the Irish Statistical Association Workshop on Frontiers in Functional Data Analysis and the Forecasting Financial Markets Conference. Han Lin Shang acknowledges the financial support of a research grant from the College of Business and Economics at the Australian National University.

}  \fi

\newpage
\bibliography{oil_FTS}
\bibliographystyle{agsm}

\newpage
\section*{Appendix A: Model Confidence Set}

\setcounter{equation}{0}
\renewcommand{\theequation}{A.\arabic{equation}}

To examine statistical significance among the three methods, we consider the Model Confidence Set (MCS). The MCS procedure proposed by \cite{HLN11} consists of a sequence of tests which permits to construct a set of ``superior" model, where the null hypothesis of equal predictive ability (EPA) is not rejected at a certain confidence level. As the EPA test statistic can be evaluated for any loss function, we adopt the tractable absolute error measure.

The procedure begins with an initial set of models of dimension $m=3$ encompassing all the model considered, $M_0 = \{M_1, M_2, M_3\}$. For a given confidence level, a smaller set, the superior set of models $\widehat{M}_{1-\alpha}^*$ is determined where $m^*\leq m$. The best scenario is when the final set consists of a single model, i. e., $m=1$. Let $d_{ij,t}$ denotes the loss differential between two models $i$ and $j$, that is
\begin{equation*}
d_{ij,t} = l_{i,t} - l_{j,t}, \qquad i,j=1,\dots,m, \quad t=1,\dots,n,
\end{equation*}
and calculate
\begin{equation*}
d_{i\cdot,t} = \frac{1}{m-1}\sum_{j\in \text{M}}d_{ij,t}, \qquad i = 1,\dots,m
\end{equation*}
as the loss of model $i$ relative to any other model $j$ at time point $t$. The EPA hypothesis for a given set of $\text{M}$ candidate models can be formulated in two ways:
\begin{align}
\text{H}_{\text{0,M}}: c_{ij}&=0, \qquad \text{for all}\quad i, j = 1,2,\dots,m\notag\\
\text{H}_{\text{A,M}}: c_{ij}&\neq 0, \qquad \text{for some}\quad i, j = 1,2,\dots,m.\label{eq:hypo_1}
\end{align}
or
\begin{align}
\text{H}_{\text{0,M}}: c_{i.}&=0, \qquad \text{for all}\quad i, j = 1,2,\dots,m\notag\\
\text{H}_{\text{A,M}}: c_{i.}&\neq 0, \qquad \text{for some}\quad i, j = 1,2,\dots,m.\label{eq:hypo_2}
\end{align}
where $c_{ij} = \text{E}(d_{ij})$ and $c_{i.} = \text{E}(d_{i.})$ are assumed to be finite and time independent. Based on $c_{ij}$ or $c_{i.}$, we construct two hypothesis tests as follows:
\begin{equation}
t_{ij} = \frac{\overline{d}_{ij}}{\sqrt{\widehat{\text{Var}}\left(\overline{d}_{ij}\right)}}, \qquad
t_{i.} = \frac{\overline{d}_{i.}}{\sqrt{\widehat{\text{Var}}\left(\overline{d}_{i.}\right)}}, \label{eq:t2} 
\end{equation}
where $\overline{d}_{i.} = \frac{1}{m-1}\sum_{j\in \text{M}}\overline{d}_{ij}$ is the sample loss of $i^{\text{th}}$ model compared to the averaged loss across models in the set $M$, and $\overline{d}_{ij} =\frac{1}{n}\sum^n_{t=1}d_{ij,t}$ measures the relative sample loss between the $i^{\text{th}}$ and $j^{\text{th}}$ models. Note that $\widehat{\text{Var}}\left(\overline{d}_{i.}\right)$ and $\widehat{\text{Var}}\left(\overline{d}_{ij}\right)$ are the bootstrapped estimates of $\text{Var}\left(\overline{d}_{i.}\right)$ and $\text{Var}\left(\overline{d}_{ij}\right)$, respectively. According to \cite{HLN11} and \cite{BC15}, we perform a block bootstrap procedure with 5,000 bootstrap samples, where the block length $p$ is given by the maximum number of significant parameters obtained by fitting an AR($p$) process on all the $d_{ij}$ term. For both hypotheses in~\eqref{eq:hypo_1} and~\eqref{eq:hypo_2}, there exist two test statistics:
\begin{equation}
T_{\text{R,M}} = \max_{i,j\in \text{M}}|t_{ij}|,\qquad T_{\max, \text{M}} = \max_{i\in \text{M}} t_{i}, \label{eq:MCS_1}
\end{equation} 
where $t_{ij}$ and $t_{i.}$ are defined in~\eqref{eq:t2}.
 
The MCS procedure is a sequential testing procedure, which eliminates the worst model at each step until the hypothesis of equal predictive ability is accepted for all the models belonging to a set of superior models. The selection of the worst model is determined by an elimination rule that is consistent with the test statistic,
\begin{equation*}
e_{\text{R,M}} = \argmax_{i\in M}\left\{\sup_{j\in M}\frac{\overline{d}_{ij}}{\sqrt{\widehat{\text{Var}}\left(\overline{d}_{ij}\right)}}\right\},\qquad e_{\max, \text{M}} = \argmax_{i\in M}\frac{\overline{d}_{i.}}{\widehat{\text{Var}}\left(\overline{d}_{i.}\right)}.
\end{equation*}

To summarise, the MCS procedure to obtain a superior set of models consists of the following steps:
\begin{enumerate}
\item[1)] Set $M=M_0$;
\item[2)] If the null hypothesis is accepted, then $M^* = M$; otherwise use the elimination rules defined in~\eqref{eq:MCS_1} to determine the worst model;
\item[3)] Remove the worst model, and go to Step 2).
\end{enumerate}

\end{document}